\documentclass[12pt,preprint]{aastex}
\begin{document}
\title{Estimating the Turn-Around Radii of Six Isolated Galaxy Groups in the Local Universe}
\author{Jounghun Lee}
\affil{Astronomy Program, Department of Physics and Astronomy, Seoul National University, 
Seoul 08826, Republic of Korea} 
\email{jounghun@astro.snu.ac.kr}
\begin{abstract}
The estimates of the turn-around radii of six isolated galaxy groups in the nearby universe are presented.  
From the Tenth Data Release of the Sloan Digital Sky Survey, we first select those isolated galaxy groups at redshifts
$z\le 0.05$ in the mass range of [$0.3$-$1$]$\times10^{14}\,h^{-1}M_{\odot}$ whose nearest neighbor groups are located at distances
larger than fiften times their virial radii. Then, we search for a gravitationally interacting web-like structure around each isolated group, 
which appears as an inclined streak pattern in the anisotropic spatial distribution of the neighbor field galaxies .  
Out of $59$ isolated groups, only seven are found to possess such web-like structures in their neighbor zones, but one of them 
turns out to be NGC 5353/4, whose turn-around radius was already measured in the previous work and thus excluded from our analysis.  
Applying the Turn-around Radius Estimator algorithm devised by Lee et al. to the identified web-like structures of the remaining six 
target groups, we determine their turn-around radii and show that three out of the six targets have larger turn-around radii than 
the spherical bound limit predicted by the Planck cosmology. We discuss possible sources of the apparent violations 
of the three groups, including the underestimated spherical bound-limit due to the approximation of the turn-around mass 
by the virial mass.
\end{abstract}

\section{Introduction}\label{sec:intro}

When a dark matter (DM) halo forms in the universe,  its gravitational struggle against the Hubble expansion in its linear stage 
leaves behind a unique vestige that is hardly effaced by any nonlinear complications in the subsequent evolution. This vestige, 
called the turn-around radius, reflects the very moment when the radial velocity of a proto-halo begins to change its sign 
as its self-gravity catches up with the Hubble flow. 
The merit and power of this vestige resides in the fact that even though it is a local quantity, it can be sufficiently well modeled by the 
linear physics. If the turn-around radii of DM halos be directly estimated from observations, then the comparison of the estimated 
values with the model predictions would put a new constraint on the initial conditions of the universe \citep{PT14,pav-etal14}. 

The linear physics predicts that in an accelerating phase of the universe the turn-around radii, $r_{t}$, of DM halos are 
bounded by a finite upper limit, $r_{t,u}$, whose value sensitively depends on the amount and equation of state of dark energy 
(DE) \citep{PT14}. 
In the standard picture where DE is given as the cosmological constant $\Lambda$ and DM is cold (i.e, $\Lambda$CDM 
cosmology),  the upper bound limit of the turn-around radii is given as 
\begin{equation}
\label{eqn:rt_u}
r_{t,u} = f\left(\frac{GM}{\Omega_{\Lambda}H^{2}}\right)^{1/3}\, ,
\end{equation}
where $H$ is the Hubble parameter, $\Omega_{\Lambda}$ is the density parameter of $\Lambda$, $M$ is the turn-around 
mass (i.e., the mass of a spherical region enclosed by the turn-around radii $r_{t}$) and $f$ is a parameter introduced 
to account for the effect of asymmetry in the DM distribution of a halo \citep{PT14}. 

According to the linear physics which precludes an individual halo from having $r_{t}\ge r_{t,u}$ (i.e., bound-limit violation) in the 
$\Lambda$CDM cosmology,  $f$ has an exact value of unity only if a DM halo forms through spherical collapse process \citep{PT14}. 
For the more realistic case of non-sperhcal DM distributions, the value $f$ has been estimated to be approximately $1.5$ on 
the whole mass scale \citep[see Figure 1 in][]{PT14}. This value of $f\approx 1.5$, however, is the empirical value obtained from 
the numerical simulations unlike the case of the spherical symmetry. 
Given that the degree of the deviation of the DM distribution from the spherical symmetry varies from halos to halos, the parameter 
$f$ for the non-spherical bound limit should be regarded as a stochastic variable and the empirically obtained value of $1.5$ as an 
average of this stochastic variable. Hence, the stochastic nature of the parameter $f$ implies that it should be possible 
for individual halos with non-spherical DM distributions to violate the {\it spherical} bound limit on rare occasions since the 
spherical bound limit is $1.5$ times lower on average than the non-spherical counterpart. 

\citet{LL17} numerically examined whether or not this theoretical prediction derived purely from the linear physics is valid in the 
deeply nonlinear regime. Directly measuring the mean turn-around radii averaged over sample halos on various mass scales from the 
MultiDark Planck simulations \citep{mdark},  they confirmed that the mean turn-around radius never exceeds both of the spherical and 
the non-spherical bound limits (say, $r^{(s)}_{t,u}$ and $r^{(ns)}_{t,u}$, respectively) on the whole mass scale.  
\citet{LL17} also explored the probability of finding an individual DM halo with $r_{t}\ge r^{(s)}_{t,u}$ and found that $14\%$ of DM halos with 
masses equal to or larger than $10^{13}\,h^{-1}M_{\odot}$ violate the spherical bound limit in a Planck cosmology \citep{planck14}.  
Besides, their analysis revealed that in a modified gravity (MG) model where the notion of DE is replaced by the deviation of the gravitational law 
from the general relativity \citep[e.g.,][and references therein]{vainshtein_review}, the frequency of the occurrences of the spherical 
bound limit violations becomes elevated, which indicated that it should be in principle possible to test the gravitational law on the cosmological 
scale by exploring the rareness of the spherical bound limit violations. 

As mentioned in \citet{PT14}, the optimal targets for the observational inspection of the bound limit violations are not the clusters 
but the groups of galaxies, especially those located in the low-density environment. It is because Equation (\ref{eqn:rt_u}) was derived 
under the assumption that a DM halo already reaches a complete relaxation state, which can be hardly justified for the case of the galaxy 
clusters.  Despite that the detections of a couple of bound violating cluster and supercluster were reported by the previous works 
\citep[e.g.,][]{kar-etal14,pea-etal14}, the detections did not attract much attention, as it was suspected that the apparent violations of the spherical bound limit by those cluster and supercluster could be ascribed to the deviation of their dynamical states from complete relaxation and/or large 
uncertainties associated with the estimates of their turn-around radii from the peculiar velocities of their neighbor galaxies.

A practical difficulty to detect the bound-limit violations on the group scale stems from the following fact: Since the galaxy groups have much 
weaker gravitational influences on their neighbor galaxies than the galaxy clusters, the conventional methodology based on the 
direct measurements of the peculiar velocity profile \citep[e.g., see][]{kar-etal14} is likely to fail in properly estimating the turn-around radii 
of galaxy groups.  Recently, \citet{lee-etal15b} developed an efficient practical algorithm dubbed  the Turn-around Radius Estimator (TRE) 
which does away with the measurements of the peculiar velocities unlike the conventional methodology. The applicability of the TRE, 
however, is contingent upon the existence of gravitationally interacting filament or sheet-like structure around a target object.

\citet{lee-etal15b} applied the TRE to NGC 5353/4, a nearby isolated galaxy group around which a thin straight filamentary structure 
had already been detected \citep{kim-etal16}, and showed that the turn-around radius of NGC 5353/4 seemed to exceed the spherical bound 
limit set by the Planck $\Lambda$CDM cosmology \citep{planck14}. This was the first observational detection of an occurrence of 
the spherical bound-limit violation on the galaxy group scale. 
A detection of a single group which appears to violate Equation (\ref{eqn:rt_u}) with $f=1$, however, cannot shatter down the $\Lambda$CDM 
cosmology, due to the stochastic nature of the parameter $f$, as mentioned in the above.  It is necessary to apply the TRE to a larger sample of 
galaxy groups and to statistically and systematically explore the frequency of the occurrences of the spherical bound limit violations, which we 
attempt to conduct in this paper.

The upcoming sections will present the followings: a concise review of the TRE in Section \ref{sec:tre_review}; the estimates of the 
turn-around radii of the nearby isolated galaxy groups via the TRE and the frequency of the occurrences of the bound limit 
violation on the group scale in Section \ref{sec:tre_estimate}; a summary of the results and the discussions of the physical implications in 
Section \ref{sec:dis}.

\section{A Brief Review of the TRE Algorithm}\label{sec:tre_review}

The TRE algorithm is based on the numerical discovery of \citet{falco-etal14} that the following universal formula depicts 
well the radial velocity profile, $v_{r}(r)$, of DM particles in the bound zone around a halo with virial radius $r_{v}$:
\begin{equation}
\label{eqn:vr}
\frac{v_{r}(r)}{H} =  r- a\frac{V_{v}}{H}\left(\frac{r}{r_{v}}\right)^{-b}\, ,
\end{equation}
where the bound-zone corresponds to the radial distances from the halo center, $r$, in the range of $(3-8)r_{v}$, $H$ is the Hubble 
parameter, and $V_{v}$ is the central velocity at $r_{v}$. \citet{falco-etal14} determined empirically the amplitude and slope parameters, 
$a$ and $b$, of Equation (\ref{eqn:vr}) with the help of a N-body simulation and suggested that the best-fit values of $a$ and $b$ 
should be universal, independent of mass scales and redshifts.  According to their claim, once the amplitude and slope parameters are 
set at the universal values, it is possible to estimate the value of $r_{v}$ (or equivalently, virial mass $M_{v}$) of a given halo by adjusting 
Equation (\ref{eqn:vr}) to the observed radial velocity profile. 

Pointing out that neither $v_{r}$ nor $r$ in Equation (\ref{eqn:vr}) is directly observable, \citet{falco-etal14} brought up the following heutristic 
scheme via which Equation (\ref{eqn:vr}) could be put into practice. Provided that a target halo is surrounded by a web-like structure 
(i.e., either a filament-like or a sheet like structure) of DM in its bound zone, it is possible to express Equation (\ref{eqn:vr}) in terms of 
the observables:
\begin{equation}
\label{eqn:vr2d}
l_{z} =  \frac{r_{2d}}{\tan\beta}- a\cos\beta\frac{V_{v}}{H}\left(\frac{r_{2d}}{\sin\beta\, r_{v}}\right)^{-b}\, ,
\end{equation}
where $l_{z}\equiv cz/H$ with relative redshift $z$ and speed of light $c$, $r_{2d}$ is the radial separation distance from the halo center 
in the projected plane of sky orthogonal to the sightline toward the target halo and $\beta$ is the angle between the position vector from 
the halo center and the sightline.  Given that both of $r_{2d}$ and $z$ are readily observable, it can be said that Equation (\ref{eqn:vr2d}) is 
a practical version of Equation (\ref{eqn:vr2d}) which has an additional parameter $\beta$ as a trade-off.  
In other words, when Equation (\ref{eqn:vr2d}) is fitted to the observed radial velocity profile by adjusting the value of $M_{v}$, 
the angle $\beta$ takes on a nuisance parameter, whose presence would unavoidably enlarge the associated statistical errors 
on $M_{v}$ \citep[see also][]{lee-etal15a}. 

Having a practical version of Equation (\ref{eqn:vr}), however, does not ensure a success in its application to real  observational data. 
It was necessary to test whether or not the radial velocity profile obtained not from DM particles but from luminous galaxies in the bound zone 
through anisotropic averaging would be well described by the same universal formula as Equation (\ref{eqn:vr2d}).  In addition, it was also 
necessary to prove the claim of \citep{falco-etal14} about the universality of the slope and amplitude parameters, $a$ and $b$, in 
Equations (\ref{eqn:vr}) and (\ref{eqn:vr2d}).

Several numerical works that were conducted in light of \citet{falco-etal14} consolidated the usefulness of 
Equations (\ref{eqn:vr}) and (\ref{eqn:vr2d}) by using larger samples from higher-resolution N-body simulations. 
For instance, \citet{lee16} proved by analyzing the data from the Millennium Run II simulations \citep{mill2} that even when the 
radial velocity profile  was obtained not from DM particles but from the galaxy-size halos, the same analytic formula as 
Equation (\ref{eqn:vr}) still validly described the numerical result. Yet, they noted that the best-fit values of $a$ and $b$ do not 
show universal constancy but exhibit variance from halos to halos, implying that not only $\beta$ but also $a$ and $b$ should be 
treated as nuisance parameters in Equation (\ref{eqn:vr2d}). 
\citet{LY16} confirmed by analyzing the data from the Multidark Planck simulations \citep{mdark} that Equation (\ref{eqn:vr}) worked even 
when the bound-zone radial velocity profile was constructed not through the isotropic averaging but through the anisotropic averaging over the 
filaments or sheets. Very recently, \citet{albaek-etal17} showed that the baryonic processes would not alter severely the functional form of the 
bound-zone radial velocity profiles. 

The key idea of \citet{lee-etal15b} who devised the TRE algorithm is that Equation (\ref{eqn:vr}) can be used to mimic  
the expansion of a proto-halo until the turn-around moment.  A proto-halo expands at a slower rate than the Hubble flow
due to its self-gravity before the turn-around moment ($t_{t}$).  Claiming that the radial velocity profile of a proto-group 
before $t_{t}$ may be well described by  the same formula as Equation (\ref{eqn:vr}),  \citet{lee-etal15b} suggested that 
Equation (\ref{eqn:vr}) should become equal to zero at the turn-around radius ($r_{t}$):
\begin{equation}
\label{eqn:rt}
r_{t} = a\frac{V_{v}}{H}\left(\frac{r_{t}}{r_{v}}\right)^{-b}\, .
\end{equation}
The procedure to estimate the value of $r_{t}$ of a massive object via the TRE algorithm can be summarized as follows 
\citep[for a detailed description, see][]{lee-etal15b}: 
(1) For a galaxy group or cluster whose viral mass $M_{v}$ is already known from priors, search for a 
filament-like or sheet-like (collectively called, web-like) structure in its neighbor zone; (2) Construct the radial velocity profile 
along the identified web-like structure; (3)  Fit the constructed radial velocity profile to Equation (\ref{eqn:vr2d}) by 
adjusting the values of $a$, $b$ and $\beta$: (4) Put the best-fit values of $a$ and $b$ into Equation (\ref{eqn:rt}) and 
find a solution to it. 
In Section \ref{sec:tre_estimate}, we will refine further this TRE algorithm and present a useful formula for the evaluation of the marginalized
errors on the estimates of the turn-around radii. 

\section{Turn-Around Radii of the Sloan Galaxy Groups}\label{sec:tre_estimate}

\citet{tempel-etal14} applied a redshift-space adapted version of the friends-of-friends (FoF) algorithm to the galaxy sample from the 
Tenth Data Release of the Sloan Digital Sky Survey (SDSS DR10) \citep{sdssdr10} to obtain a catalog of the FoF groups. 
From the catalog, one can draw out information on various properties of the FoF groups including the redshifts ($z$), equatorial 
coordinates of their centers ($RA$ and $DEC$), and their virial radii ($r_{v}$) and masses ($M_{v}$) which were estimated under the 
assumption that their DM density profiles follow the Navvaro-Frenk-White (NFW) formula \citep{nfw97}.  
They also provided the galaxy catalog from which information on the spectroscopic properties of the member galaxies belonging to each 
FoF group can be extracted.  

Analyzing the group catalog of \citet{tempel-etal14}, we select those FoF groups with $0.3\le M_{v}/(10^{14}h^{-1}M_{\odot})\le 1$ 
(typical group scale) and $z\le 0.05$, being isolated enough to be separated by their nearest groups of comparable masses by more than 
$15r_{v}$.  We consider only the isolated groups, given the  numerical result of \citet{LY16} that the best agreement between 
Equation (\ref{eqn:vr}) and the reconstructed radial velocity profile is achieved for the case of the isolated halos. 
Furthermore, the TRE algorithm substitutes $M_{v}$ for the turn-around mass in Equation (\ref{eqn:rt_u}), 
which approximation works best for the case of an isolated object.  

Selected are a total of $59$ isolated galaxy groups, in the neighbor zones of which we attempt to identify web-like structures composed 
of the field galaxies.  Although in the previous works of \citet{falco-etal14} and \citet{lee17},  the neighbor zone around a cluster was defined to 
have a large extent of $\vert l_{z}\vert \le 40\,h^{-1}$Mpc and $4\le r_{2d}/(h^{-1}{\rm Mpc})\le 20$,  we confine the neighbor zone to a much 
smaller extent of  $\vert l_{z}\vert \equiv \vert cz/H\vert\le 20\,h^{-1}$Mpc and $2\le r_{2d}/(h^{-1}{\rm Mpc})\le 10$, given that the 
target objects are not the clusters but the less massive groups whose gravitational influences can reach out only this small extent. 
It is also worth explaining here why we identify a web-like structure from the distributions only of the {\it field} galaxies, excluding the wall 
galaxies. It is because the wall galaxies unlike the field counterparts are expected to be heavily influenced by their own hosts even if they are 
located in the same neighbor zone.  

\begin{figure}
\begin{center}
\plotone{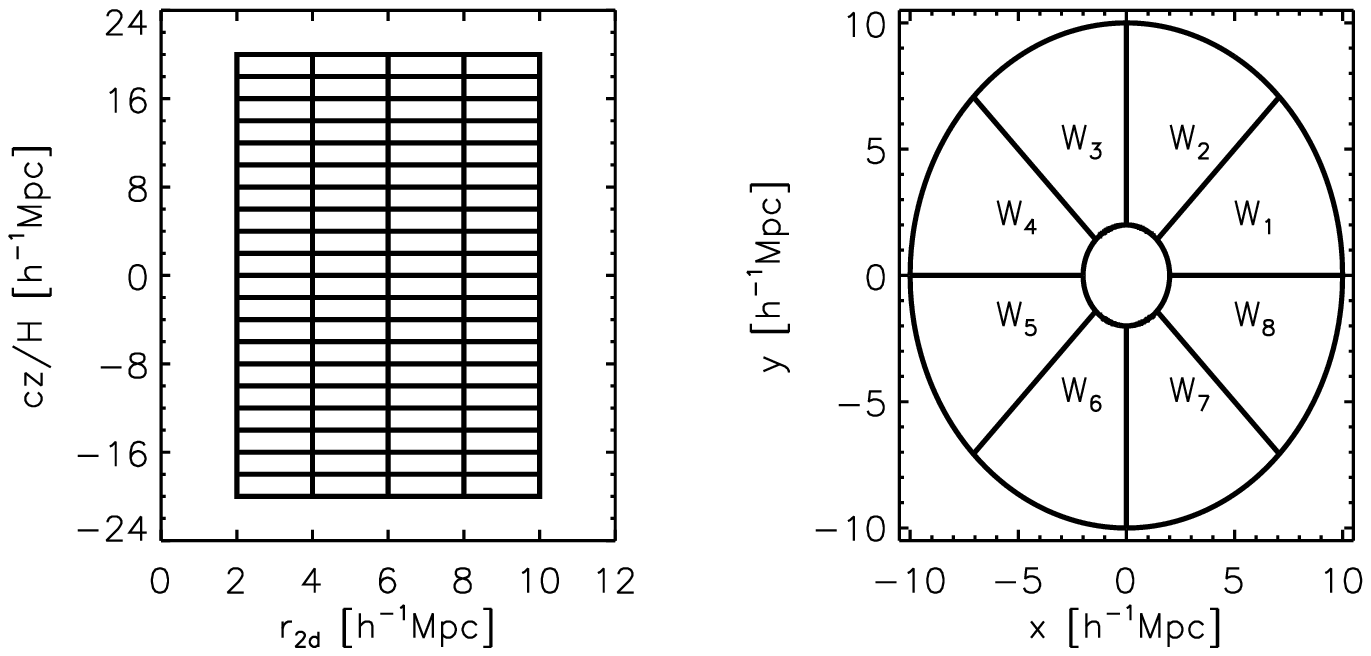}
\caption{(Left panel:) Illustration of the pixelation of the bound zone around an isolated group in two
dimensional space spanned by $r_{2d}$ and $cz/H$, where $r_{2d}$ is the separation distance from the group center 
in the plane of sky perpendicular to the line of sight toward the group. (Right Panel:)
Illustration of the division of the plane of the sky around each isolated group into eight wedges of 
equal area.}
\label{fig:linepie}
\end{center}
\end{figure}

Adopting the methodology suggested by \citet{falco-etal14} for the identification of the web-like structures, we first look for the overdense pixels in 
the neighbor zone around each isolated group by counting the field galaxies. The neighbor zone around each isolated group is partitioned into $80$ 
pixels of equal sizes in two dimensional space spanned by $r_{2d}$ and $l_{z}$, as illustrated in the left panel of Figure \ref{fig:linepie}.  
The spherical shell, with inner and outer radii of $2\,h^{-1}$Mpc and $10\,h^{-1}$Mpc, respectively, are also partitioned into eight wedges 
(say, $\{W_{i}\}_{i=1}^{8}$), as depicted in the right panel of Figure \ref{fig:linepie}, where $(x,y)$ denotes a two dimensional position vector 
from the group center in the equatorial coordinate system, with $r_{2d}=(x^{2}+y^{2})^{1/2}$. 
Each wedge represents a realization of the neighbor zone of a given isolated group, and the eight wedges form an ensemble over which the 
average residual number densities of the neighbor field galaxies around the group will be evaluated.

From the galaxy sample from the SDSS DR10, we select those field galaxies which belong to the neighbor zone of each isolated 
group by estimating the values of $r_{2d}$ and $l_{z}$.  Then, we investigate to which pixel and to which wedge each of the 
neighbor field galaxies belong.  Suppose that one wants to find the dimensionless residual number density of the neighbor 
field galaxies at the $ij$ th pixel of the wedge $W_{1}$ (say, $\delta_{ij}^{W_1}$). The first step is to compute the number densities of 
the neighbor field galaxies belonging to the $ij$th pixel of the wedge $W_{1}$. The second step is to compute the number densities 
of the neighbor field galaxies at the same pixel of five different wedges, $W_{3},\ W_{4},\ W_{5},\ W_{6},\ W_{7}$. 
Two wedges, $W_{2}$ and $W_{8}$, adjacent to the wedge $W_{1}$ (see the right panel of Figure \ref{fig:linepie}) are excluded. 
The third step is to compute the back ground number density of the neighbor field galaxies at the $ij$th pixel by taking the average over 
the five wedges.  

The fourth step is to evaluate the residual number density at the $ij$th pixel of the wedge $W_{1}$, $\delta_{ij}^{1}$, by dividing the 
difference in the results between the first and the third steps by the background number density at the $ij$th pixel. 
The fifth step is to compute the standard deviation, $\sigma_{ij}^{W_1}$, of the residual number density, in a similar manner. 
The final step is to see whether or not the condition of $\delta_{ij}^{W_1}\ge \sigma_{ij}^{W_1}$ is met at the $ij$th pixel of the 
wedge $W_{1}$.  If met, the $ij$th pixel of the wedge $W_{1}$ is selected as a candidate overdense site where a web-like structure 
composed of the neighbor field galaxies may be found.  Retake this procedure repeatedly for the other pixels and wedges to find all 
the overdense sites in the neighbor zone of each isolated group. See \citet{falco-etal14} for a detailed description. 

\begin{figure}
\begin{center}
\plotone{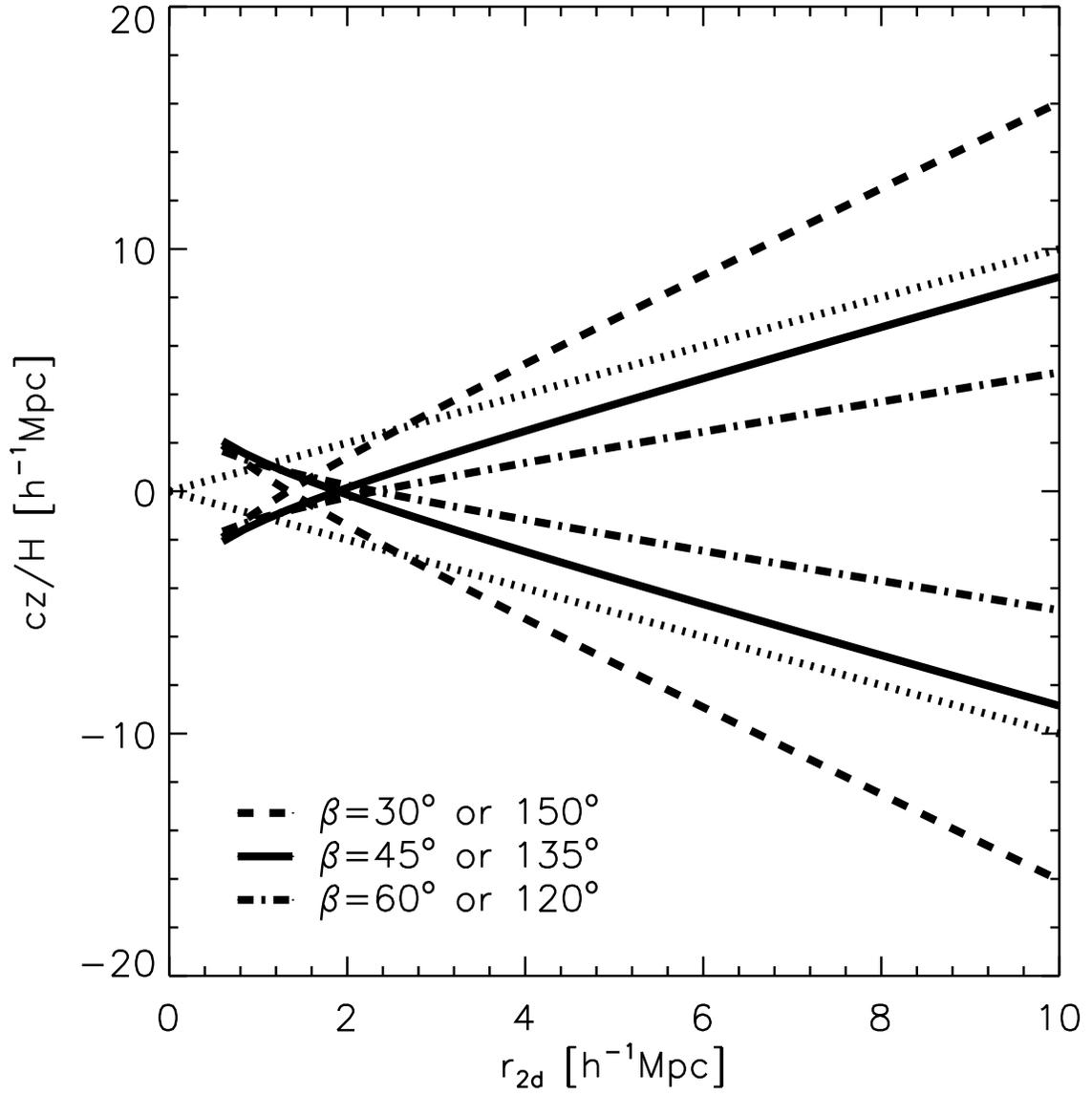}
\caption{Projected radial velocity profiles of the bound-zone galaxies around a galaxy group with 
viral mass of $5\times 10^{13}\,h^{-1}M_{\odot}$ for six different cases of the inclination angle, $\beta$.}
\label{fig:test_vr}
\end{center}
\end{figure}

Before proceeding to identify a web-like structure in the overdense pixels of the neighbor zone around each isolated group, it is 
worth emphasizing that the TRE algorithm would be applicable only to those pixels which would appear altogether as inclined streak 
lines in the $r_{2d}$-$l_{z}$ configuration space, as explained in \citet{bri-etal16}. 
Figure \ref{fig:test_vr} plots the analytic formula of Equation (\ref{eqn:vr2d}) for six different cases of $\beta$, setting $M_{v}$ at 
$5\times 10^{13}\,h^{-1}M_{\odot}$. We look for a web-like structure composed of the field galaxies located in the neighbor zone 
around a target group, which would appear similar to the inclined lines shown in Figure \ref{fig:test_vr}. 

We find that only seven out of the $59$ isolated groups possess such web-like structures in their neighbor zones. 
Among the seven groups, one turns out to be NGC 5353/4, whose turn-around radius was already estimated by \citet{lee-etal15b} 
to exceed the spherical bound limit. NGC 5353/4 being excluded, the rest six groups (say, GG1, GG2, GG3, GG4, GG5, GG6) become 
our target groups to which the TRE algorithm is going to be applied for the estimation of their turn-around radii. Table \ref{tab:nsc} 
presents the equatorial coordinates, redshifts and virial masses of the six target groups. 

Figure \ref{fig:zr1} shows as red closed circles the locations of the neighbor field galaxies belonging to the overdense 
pixels around GG1 in the two dimensional configuration space spanned by $r_{2d}$ and $l_{z}$. The green dotted line 
correspond to the locations at which the condition of $l_{z}=r_{2d}$ is met, while the blue closed circles represent the 
configurations of the member galaxies of GG1.  Noting the existence of an inclined streak of the neighbor field galaxies in the 
overdense pixels of the wedge $W_{6}$, which look similar to the inclined lines shown in Figure \ref{fig:test_vr}, it is identified 
a web-like structure around GG1 and shown as black open circles in Figure \ref{fig:zr1}.  Figures \ref{fig:zr2}-\ref{fig:zr6} show 
the same as Figure \ref{fig:zr1} but for the other five target groups. As can be seen, the web-like structures around 
GG2, GG3, GG4, GG5, GG6 are identified in the wedges of $W_{8},\ W_{7},\ W_{4},\ W_{2},\ W_{4}$, respectively. 
Figure \ref{fig:zr7} shows the same as Figures \ref{fig:zr1}-\ref{fig:zr6} but for the case of an isolated group around 
which no web-like structure is identified and thus not selected as a target. 

Suppose that we identify a web-like structure composed of $n_{f}$ neighbor field galaxies from one of the eight wedges around a target 
group. Employing the maximum likelihood method as \citet{lee-etal15b} and \citet{lee17} did, we determine the best-fit values of 
$a,\ b,\ \beta$ in Equation (\ref{eqn:vr2d}), which maximizes the following posterior distribution: 
\begin{eqnarray}
\label{eqn:post}
p(a,b,\beta) &\propto& \exp\left[-\frac{\chi^{2}(a,b,\beta)}{2}\right]\, ,\\
\label{eqn:chi2}
\chi^{2}(a,b,\beta) &=& \sum_{k=1}^{n_{f}} \left[l_{z,k}-l_{z,k}^{T}(a,b,\beta)\right]^{2} \, ,\\
\label{eqn:lT}
l_{z,k}^{T}(a,b,\beta) &=& \frac{r_{2d,k}}{\tan\beta}- a\cos\beta\frac{V_{v}}{H}\left(\frac{r_{2d,k}}{\sin\beta\, r_{v}}\right)^{-b}\, ,
\end{eqnarray}
where $(r_{2d,k},\ l_{z,k})$ are the observed position vector of the $k$th neighbor field galaxy belonging to an identified web-like 
structure, while $l^{T}_{z,k}$ represents Equation (\ref{eqn:vr2d}) with $r_{2d}$ set at $r_{2d,k}$. 

To improve the efficiency of the TRE algorithm in its practical application, we rearrange the terms of Equation (\ref{eqn:rt}) to derive the 
following closed analytic expression for $r_{t}$ as a function of $a$ and $b$:
\begin{equation}
\label{eqn:rt2}
r_{t}(a,b) = \exp\left\{\frac{1}{(1+b)}\ln\left[r^{b}_{\rm v}\left(\frac{aV_{v}}{H}\right)\right]\right\}\, .
\end{equation}
Suppose that the posterior function, Equation (\ref{eqn:post}), is found to reach its maximum at $a=\hat{a},\ b=\hat{b},\ \beta=\hat{\beta}$. 
Putting the best-fit values, $\hat{a}$ and $\hat{b}$, into Equation (\ref{eqn:rt2}), one can readily estimate the turn-around radius of a target group 
as $\hat{r}_{t}(\hat{a},\hat{b})$.  

We estimate the associated errors on $\hat{r}_{t}$, $\sigma_{r_{t}}$, according to the error propagation formula \citep{WJ12}:
\begin{equation}
\label{eqn:sig_rt}
\sigma^{2}_{r_{t}}\approx \left(\frac{\partial r_{t}}{\partial a}\right)\Bigg{\vert}_{\hat{a},\hat{b}}^{2}\sigma^{2}_{a} + 
\left(\frac{\partial r_{t}}{\partial b}\right)\Bigg{\vert}_{\hat{a},\hat{b}}^{2}\sigma^{2}_{b}  
+ 2\left(\frac{\partial r_{t}}{\partial a}\right)\Bigg{\vert}_{\hat{a},\hat{b}}
\left(\frac{\partial r_{t}}{\partial b}\right)\Bigg{\vert}_{\hat{a},\hat{b}}{\rm cov}(a, b)\, ,
\end{equation}
where $\sigma_{a}$ and $\sigma_{b}$ denote the {\it marginalized} errors in the determination of the best-fit values of $a$ and $b$, 
respectively, and ${\rm cov}(a,b)$ is the marginalized covariance between $a$ and $b$, three of which can be calculated as
\begin{eqnarray}
\label{eqn:sig_a}
\sigma^{2}_{a} &=& \int d\beta\int da \int db~(a-\langle a\rangle)^{2}\,p(a,b,\beta)\, , \\
\label{eqn:sig_b}
\sigma^{2}_{b} &=& \int d\beta \int da \int db~(b-\langle b\rangle)^{2}\,p(a,b,\beta)\, , \\
\label{eqn:sig_c}
{\rm cov}(a, b) &=& \int d\beta \int da \int db~(a-\langle a\rangle)(b-\langle b\rangle)\,p(a,b,\beta)\,  ,
\end{eqnarray}
where $\langle a\rangle=\int d\beta \int da \int db~a\,p(a,b,\beta)$ and $\langle b\rangle=\int d\beta \int da \int db~b\,p(a,b,\beta)$. 
The best-fit values of $a$ and $b$ determined by the maximum likelihood method along with Equations (\ref{eqn:post})-(\ref{eqn:lT}) 
as well as the associated errors and covariances estimated by Equations (\ref{eqn:sig_a})-(\ref{eqn:sig_c}) for the six target groups 
are presented in Table \ref{tab:abcov}. 

We calculate the partial derivatives, $\partial r_{t}/\partial a$ and $\partial r_{t}/\partial b$ at $a=\hat{a}$ and $b=\hat{b}$, 
in Equation (\ref{eqn:sig_rt}) and derive the following expressions.
\begin{equation}
\label{eqn:partial}
\left(\frac{\partial r_{t}}{\partial a}\right)\Bigg{\vert}_{\hat{a},\hat{b}} = \frac{\hat{r}_{t}}{\hat{a}(1+\hat{b})}\, ,\qquad
\left(\frac{\partial r_{t}}{\partial b}\right)\Bigg{\vert}_{\hat{a},\hat{b}} = 
\frac{\hat{r}_{t}}{\hat{a}(1+\hat{b})}\ln\left(\frac{r_{v}}{\hat{r}_{t}}\right)\, .
\end{equation}
Through Equations (\ref{eqn:sig_rt})-(\ref{eqn:partial}), we finally estimate $\sigma_{r_t}$ for each target group. 
It is worth emphasizing here that although the error on the nuisance parameter $\beta$, $\sigma_{\beta}$, does not explicitly appear in 
Equation (\ref{eqn:sig_rt}), the variation of $\beta$ is taken into full account for the determination of the marginalized error, $\sigma_{r_{t}}$, 
since both of $\sigma_{a}$ and $\sigma_{b}$ are determined by the simultaneous marginalization of the posterior distribution, 
$p(a,b,\beta)$, over $a$,$b$ and $\beta$. 

Table \ref{tab:rt} lists the estimated turn-around radii and the associated errors for the six targets, and compare the values with the spherical 
and non-spherical bound limits set by the Planck cosmology. 
As can be seen, for the cases of GG1, GG2, and GG3, the differences between $\hat{r}$ and $r^{(s)}_{t,u}$ are larger than $\sigma_{r_{t}}$, 
while for the cases of GG4, GG5, and GG6, the differences fall within $\sigma_{r_t}$.  Given this result, the former three groups could be regarded 
as candidates for the spherical bound limit violation. 
Yet, the comparison of $\hat{r}_{t}-r^{(ns)}_{t,u}$ with $\sigma_{r_t}$ reveals that none of the six targets violate the non-spherical bound-limit. 

\section{Summary and Discussion}\label{sec:dis}

\begin{figure}
\begin{center}
\plotone{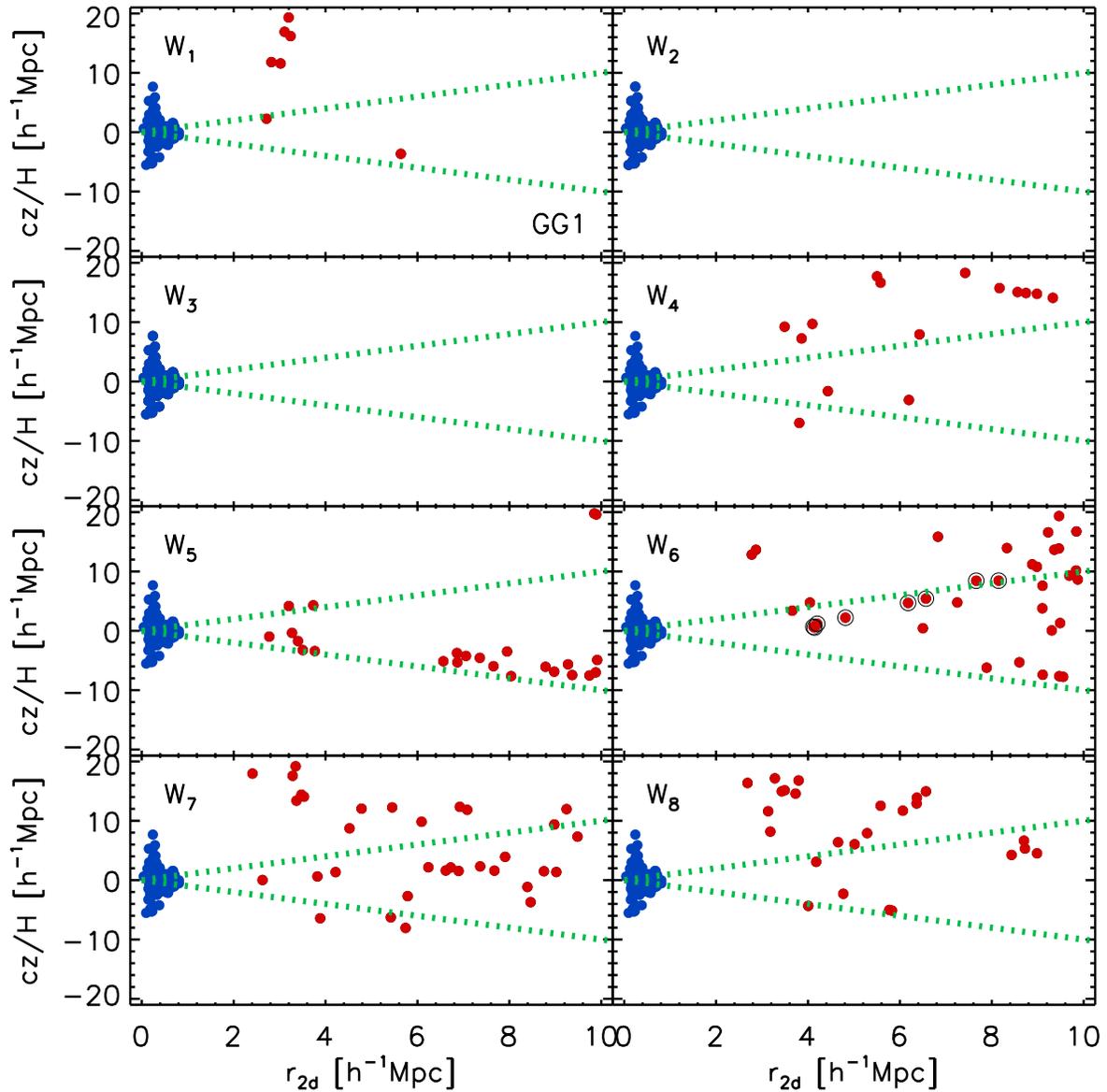}
\caption{Configurations of the field galaxies (filled red dots) located in the overdense sites around 
an isolated galaxy group, GG1 (blue dots) from the catalog of \citet{tempel-etal14}.  The open black 
dots indicate those field galaxies belonging to a web-like structure identified around GG1. }
\label{fig:zr1}
\end{center}
\end{figure}

\begin{figure}
\begin{center}
\plotone{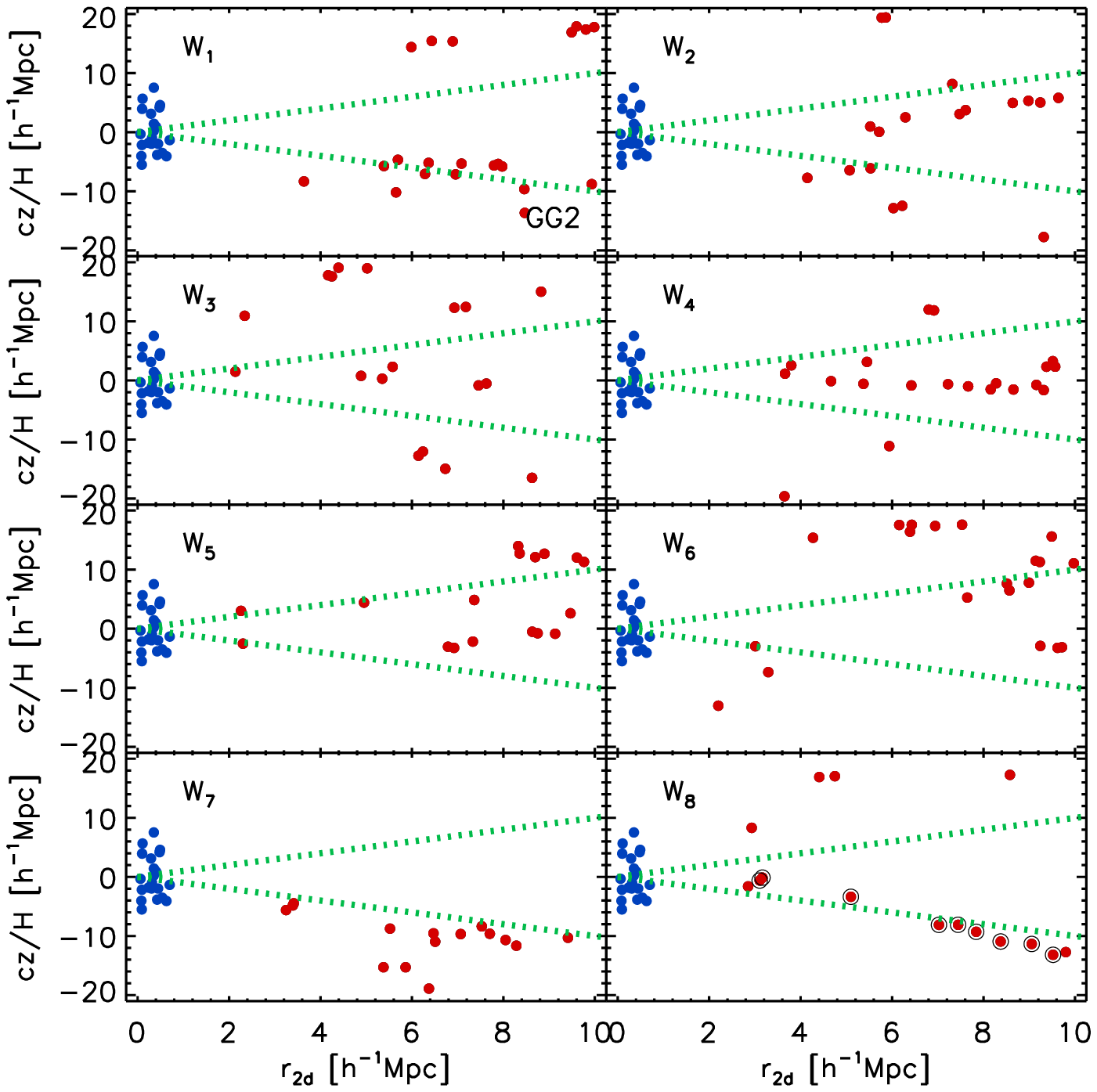}
\caption{Same as Figure \ref{fig:zr1} but for a different galaxy group, GG2.}
\label{fig:zr2}
\end{center}
\end{figure}

\begin{figure}
\begin{center}
\plotone{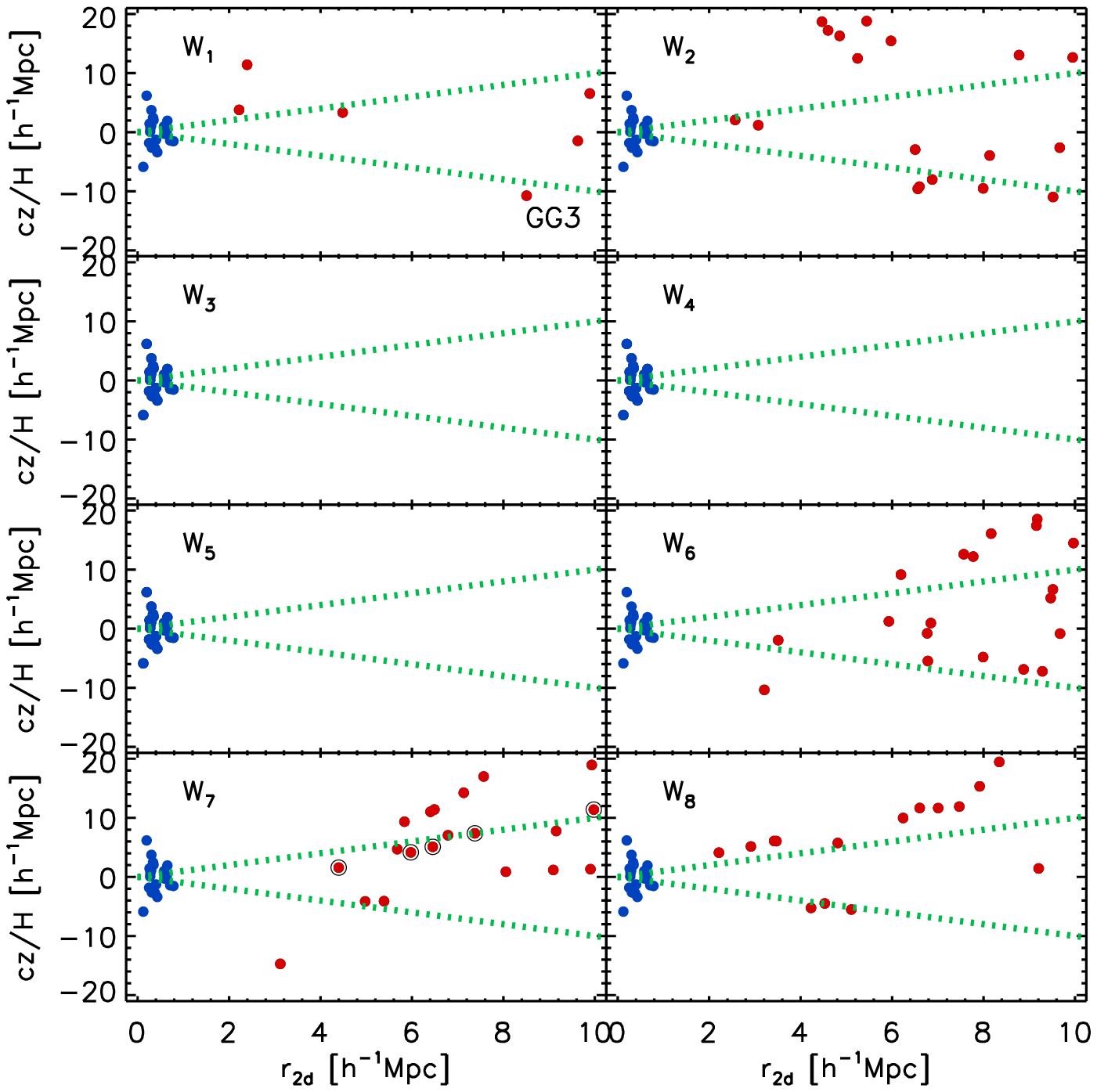}
\caption{Same as Figure \ref{fig:zr1} but for a different galaxy group, GG3.}
\label{fig:zr3}
\end{center}
\end{figure}

\begin{figure}
\begin{center}
\plotone{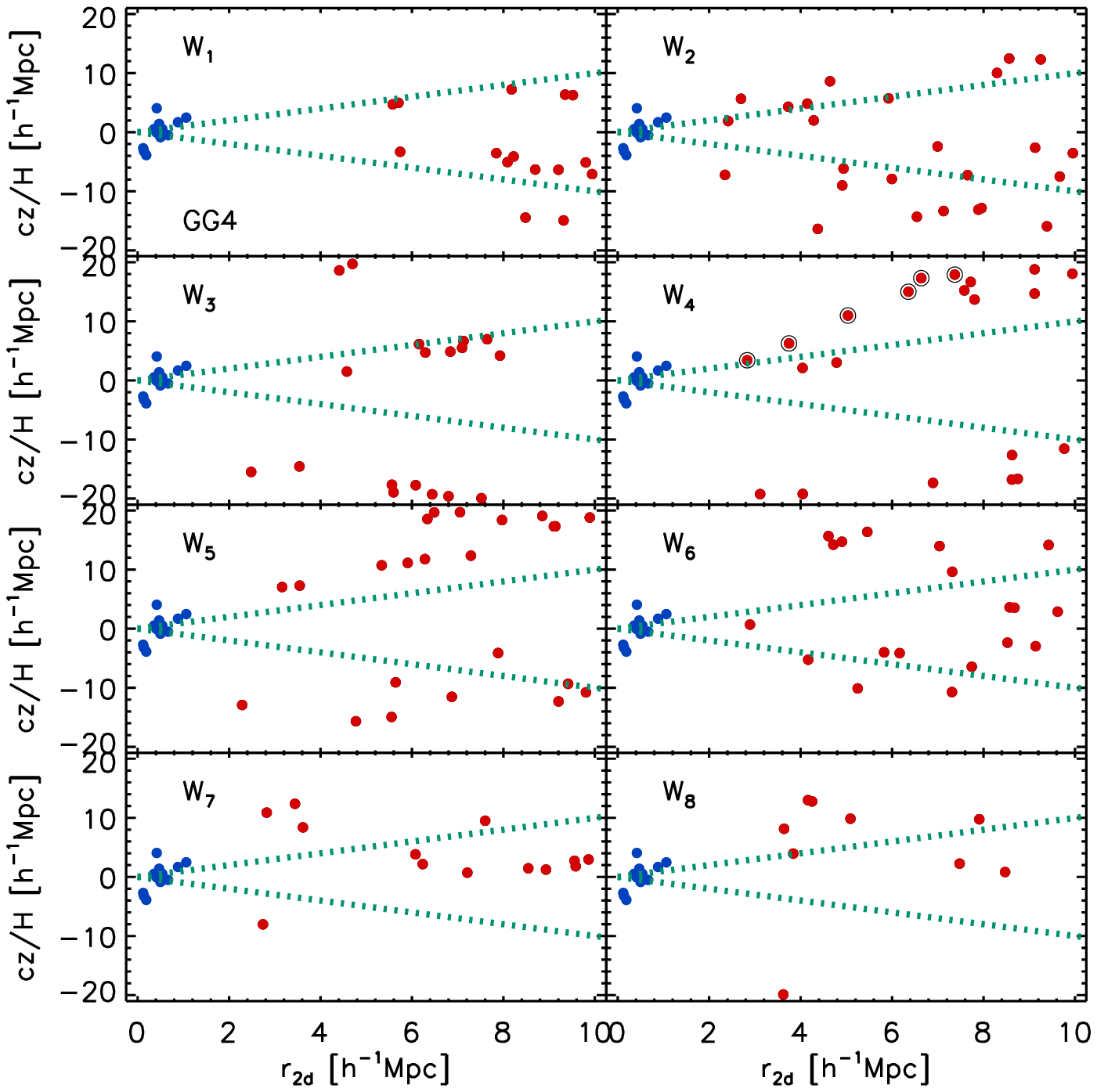}
\caption{Same as Figure \ref{fig:zr1} but for a different galaxy group, GG4.} 
\label{fig:zr4}
\end{center}
\end{figure}

\begin{figure}
\begin{center}
\plotone{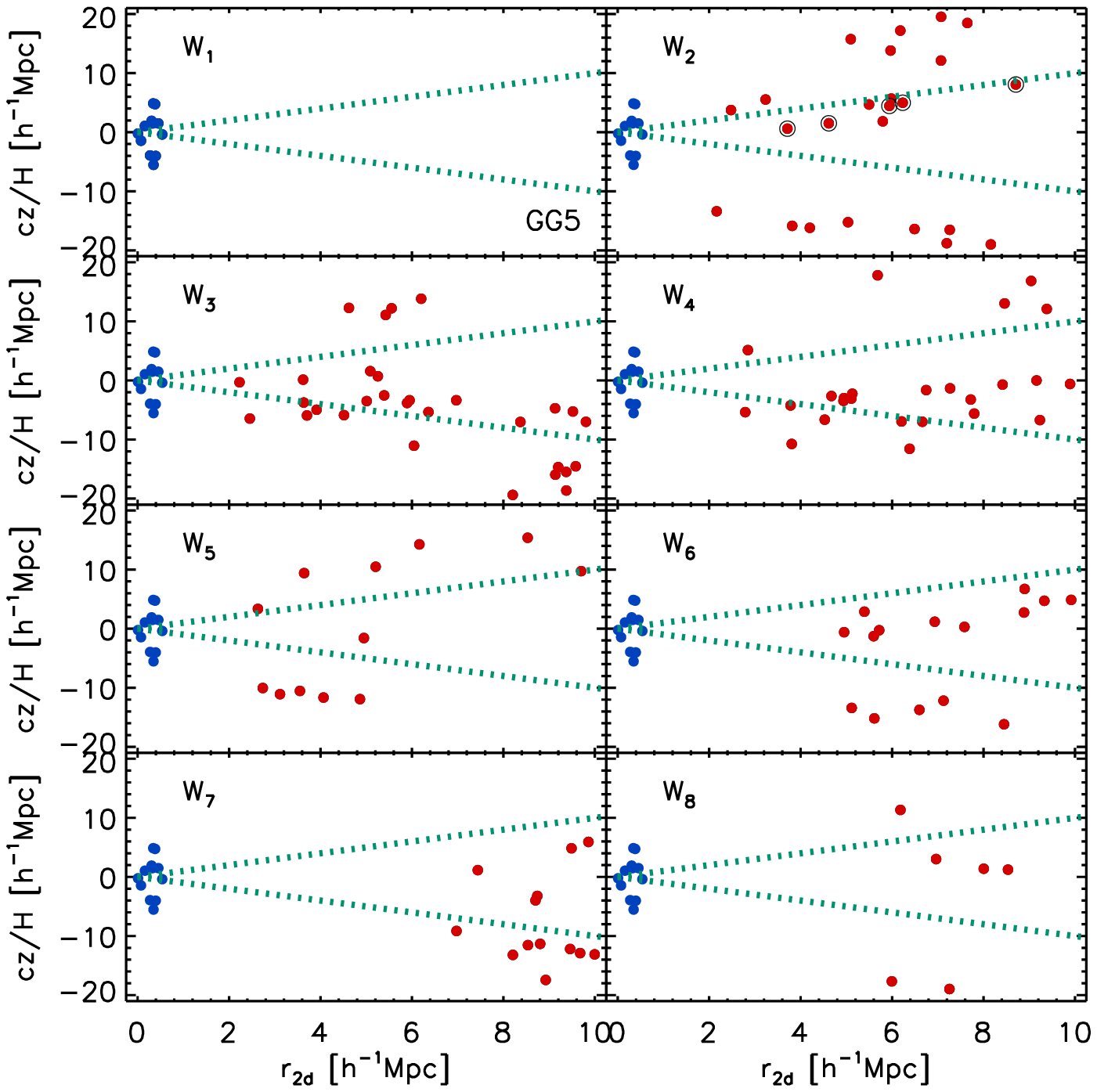}
\caption{Same as Figure \ref{fig:zr1} but for a different galaxy group, GG5.}
\label{fig:zr5}
\end{center}
\end{figure}

\begin{figure}
\begin{center}
\plotone{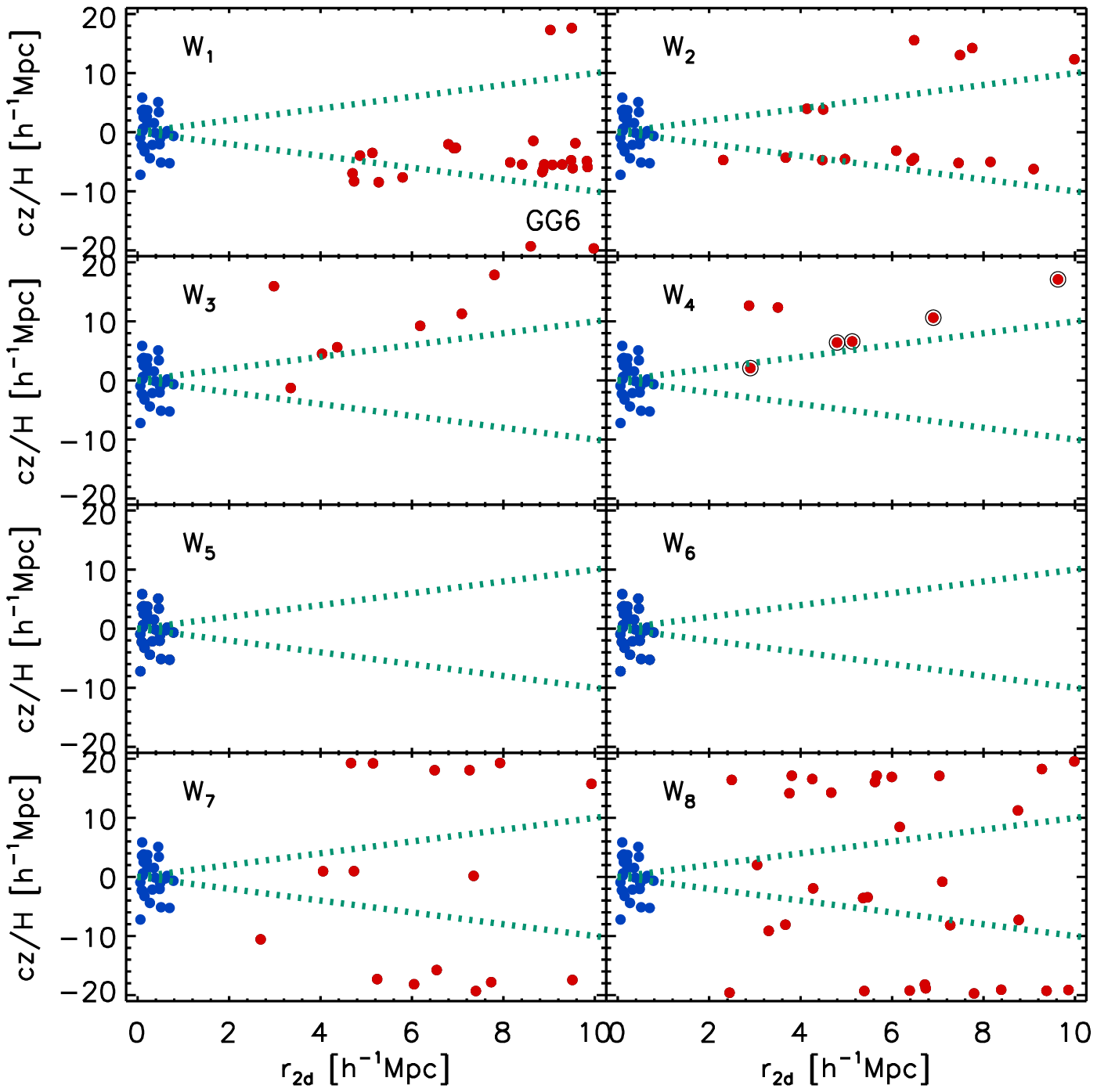}
\caption{Same as Figure \ref{fig:zr1} but for a different galaxy group, GG6.}
\label{fig:zr6}
\end{center}
\end{figure}

\begin{figure}
\begin{center}
\plotone{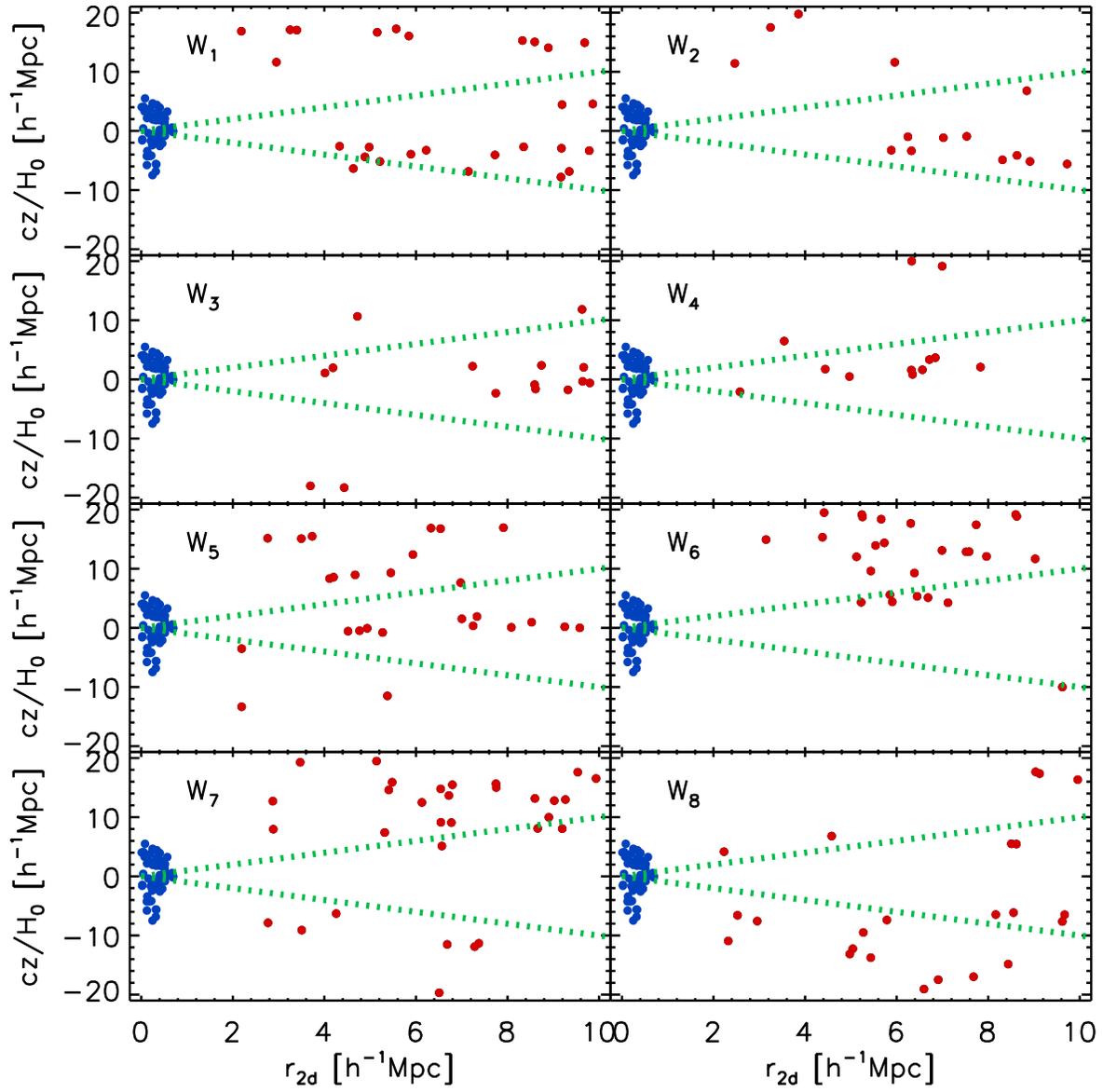}
\caption{Same as Figure \ref{fig:zr1} but for a target group in the neighbor zone of which no web-like 
structure of the field galaxies is identified. }
\label{fig:zr7}
\end{center}
\end{figure}

Employing the TRE algorithm developed by \citet{lee-etal15b}, we have estimated the turn-around radii of six isolated galaxy groups with masses 
in the range of $0.3\le M_{v}/[10^{14}h^{-1}M_{\odot}]\le 1$ at redshifts of $z\le 0.05$ from the SDSS DR10. To ensure the validity and 
efficacy of the TRE algorithm, our analysis has been restricted to the local isolated galaxy groups around which the neighbor field galaxies exhibit 
anisotropic spatial distributions.  For each of the six targets, we have constructed a radial velocity profile along the anisotropic distribution of the 
neighbor field galaxies (Figures \ref{fig:zr1}-\ref{fig:zr6}) and fitted it to the analytic formula derived by \citet{falco-etal14}. Finally, the turn-around 
radius of each target has been determined as the radial distance at which the best-fit formula hits zero, and the marginalized errors propagated 
through the fitting procedure has been also evaluated (Table \ref{tab:rt}).  

The measured turn-around radii of the six targets have been compared with the spherical and non-spherical upper bound limits predicted 
by the $\Lambda$CDM cosmology. Among the six targets, three have been shown to violate the spherical bound-limit, while the other three abide 
by it.  Although no violation of the non-spherical bound limit is found, we have noted that the observed frequency at which the spherical bound 
violation occurs on the galaxy group scale is rather high compared with the numerical result of \citet{LL17} who found the frequency as low as 
$14\%$ in a $\Lambda$CDM universe. Yet, before rushing to a conclusion that our observational result challenges the $\Lambda$CDM cosmology, 
it should be worth inspecting a more mundane source of this rather high frequency of the occurrence of the spherical bound-limit violation. 

The first suspicion falls on the underestimate of the spherical bound limit caused by substituting the virial mass for the turn-around mass in 
Equation (\ref{eqn:rt_u}).  Although it has been presumed throughout our current analysis that for the case of the isolated galaxy group the virial 
mass would approximate well the turn-around mass, it has yet to be quantitatively addressed how close the virial mass of each target is to its turn-
around mass, how the difference between the two masses would depend on the mass scale, and how significantly the difference 
would change the value of the spherical bound limit.
 
Another factor that has not been taken into account but may have contaminated the final result is  the uncertainties associated with the 
measurements of the virial masses of the galaxy groups. \citet{tempel-etal14} measured the virial masses of the SDSS groups 
under the assumption that the DM density profiles are well described by the universal NFW formula \citep{nfw97}.  However, 
several numerical experiments already invalidated the concept of the {\it universality} of the NFW density profile 
\citep[e.g.,][]{nav-etal04}. The deviation of the true density profiles from the NFW formula may have caused systematic errors in the 
measurements of the virial masses of the target groups, which may have in turn contaminated our estimates of their turn-around radii. 

The other downside is the small size of our sample consisting only of six target groups, which obstructs a statistically conclusive 
interpretation of the final result. This small sample size is an inevitable outcome of the generic limitation of the TRE algorithm which is 
applicable only to those isolated groups having web-like structures in their neighbor zones.  Furthermore, since a web-like structure had to be 
identified from the anisotropic spatial distribution of the {\it field} galaxies to guarantee its gravitational link with the target, each identified web-like 
structure has turned out have a very low richness, which incurred inaccuracy in the construction of the radial velocity profiles. Our future work 
will be in the direction of addressing these remaining issues and improving further the statistical analysis as well as the TRE algorithm. 

\acknowledgements

The manuscript has been significantly improved from the original version through revision thanks to many valuable suggestions from an 
anonymous referee. I acknowledge the support of the Basic Science Research Program through the National Research Foundation (NRF) of Korea 
funded by the Ministry of Education (NO. 2016R1D1A1A09918491).  I was also partially supported by a research grant from the 
NRF of Korea to the Center for Galaxy Evolution Research (No.2017R1A5A1070354).

\clearpage
\clearpage
\clearpage
\clearpage
\clearpage
\clearpage
\clearpage
\clearpage
\clearpage
\clearpage
\begin{deluxetable}{ccccc}
\tablewidth{0pt}
\setlength{\tabcolsep}{5mm}
\tablecaption{Equatorial coordinates, redshifts, virial masses of the target groups}
\tablehead{Group & $RA$ & $DEC$ & $z$ & $M_{v}$  \\
& ($^{\circ}$) & ($^{\circ}$) & & ($10^{12}\,h^{-1}M_{\odot}$)}
\startdata
GG1 & $226.08$  & $1.64$ & $0.007$ & $40.85$ \\
GG2 & $144.43$ & $17.06$ & $0.029$ & $60.69$\\
GG3 & $157.20$ & $8.59$ & $0.049$ & $37.00$ \\
GG4 & $226.81$ & $9.59$ & $0.045$ & $30.13$\\
GG5 & $254.37$ & $27.35$ & $0.037$ & $48.72$\\
GG6 & $123.70$ & $55.16$ & $0.033$ & $48.85$\\
\enddata
\label{tab:nsc}
\end{deluxetable}
\clearpage
\begin{deluxetable}{cccccc}
\tablewidth{0pt}
\setlength{\tabcolsep}{5mm}
\tablecaption{Best-fit parameters and their covarances for the target groups}
\tablehead{Group & $\hat{a}$ & $\sigma_{a}$ & $\hat{b}$ & $\sigma_{b}$ & cov($a$,$b$)}
\startdata
GG1 & $1.13$ & $0.32$ & $-0.13$ & $0.25$ & $0.06$\\
GG2 & $3.67$ & $1.49$ & $0.44$ &$0.16$ & $0.03$\\
GG3 & $2.37$ & $0.58$ & $0.23$ &$0.29$ & $0.16$\\
GG4 &  $3.77$ & $1.11$ & $0.43$ & $0.37$ & $0.37$ \\
GG5 &  $6.20$ & $1.61$ & $0.82$ & $0.43$ & $0.64$ \\
GG6 & $0.37$ & $0.07$ & $-0.39$ & $0.16$ & $0.01$\\
\enddata
\label{tab:abcov}
\end{deluxetable}
\clearpage
\begin{deluxetable}{ccccc}
\tablewidth{0pt}
\setlength{\tabcolsep}{5mm}
\tablecaption{Turn-around radii of the target groups and the spherical and non-spherical bound limits}
\tablehead{Group & $\hat{r}_{t}$ & $\sigma_{r_{t}}$ & $r^{(s)}_{t,u}$ & $r^{(ns)}_{t,u}$ \\
& ($h^{-1}$Mpc) & ($h^{-1}$Mpc) & ($h^{-1}$Mpc) & ($h^{-1}$Mpc)}
\startdata
GG1 & $9.01$ & $4.47$ & $3.86$ & $5.01$ \\
GG2 & $7.72$ & $2.85$ & $4.40$ & $5.72$ \\
GG3 & $6.94$ & $2.96$ & $3.73$ & $4.85$ \\
GG4 &  $6.34$ & $3.03$ & $3.49$ & $4.53$ \\
GG5 &  $5.42$ & $2.23$ & $3.95$ & $5.92$  \\
GG6 & $4.89$ & $1.26$ & $4.09$ & $5.32$ \\
\enddata
\label{tab:rt}
\end{deluxetable}

\end{document}